\def\be{\begin{equation}}
\def\ee{\end{equation}}
\def\bea{\begin{eqnarray*}}
\def\eea{\end{eqnarray*}}

\documentclass[useAMS,usenatbib]{mn2e}
\usepackage{subfigure}
\usepackage{graphicx}
\begin{document}

\title{The AzTEC mm-Wavelength Camera}


\author[G.W. Wilson et al.]{
G.W.~Wilson,$^1$ 
J.E.~Austermann,$^1$ 
T.A.~Perera,$^1$ 
K.S.~Scott,$^1$ 
P.A.R.~Ade,$^2$ 
\newauthor
J.J.~Bock,$^3$ 
J.~Glenn,$^4$ 
S.R.~Golwala,$^5$ 
S.~Kim,$^6$ 
Y.~Kang,$^6$ 
D.~Lydon,$^1$ 
\newauthor
P.D.~Mauskopf,$^2$ 
C.R.~Predmore,$^7$ 
C.M.~Roberts,$^1$ 
K.~Souccar$^1$ 
and M.S.~Yun$^1$\\
$^1$Department of Astronomy, University of Massachusetts, Amherst, MA 01003.\\
$^2$Physics and Astronomy, Cardiff University, 5, The Parade, P.O. Box 913, Cardiff CF24 3YB, Wales, UK.\\
$^3$Jet Propulsion Laboratory, California Institute of Technology, 4800 Oak Grove Drive, Pasadena, CA 91109.\\
$^4$Center for Astrophysics and Space Astronomy, 389-UCB, University of Colorado, Boulder, CO, 80309.\\
$^5$California Institute of Technology, 1200 East California Boulevard, MC 59-33, Pasadena, CA 91125.\\
$^6$Astronomy \& Space Science Department, Sejong University, 98 Kwangjin-gu, Kunja-dong, 143-747, Seoul, South Korea.\\
$^7$Predmore Associates, South Deerfield, MA 01373.\\
}

\date{\today}

\pagerange{\pageref{firstpage}--\pageref{lastpage}} \pubyear{2007}

\maketitle

\label{firstpage}

\begin{abstract}
AzTEC is a mm-wavelength bolometric camera utilizing 144 silicon
nitride micromesh detectors.  Herein we describe the AzTEC instrument
architecture and its use as an astronomical instrument.  We report on
several performance metrics measured during a three month observing
campaign at the James Clerk Maxwell Telescope, and conclude with our
plans for AzTEC as a facility instrument on the Large Millimeter
Telescope.
\end{abstract}

\begin{keywords}
instrumentation:detectors, submillimetre, galaxies:starburst, galaxies:high redshift
\end{keywords}

\section{INTRODUCTION}
\label{sec:int} 
In the era of ALMA, the scientific niche for large mm-wave telescopes
comes in the ability to use array receivers to make large surveys of
the sky.  The last decade has seen a series of major advances in
mm-wavelength bolometric instruments and it is now commonplace for
ground-based telescopes to have mm/submm wavelength cameras with tens
to hundreds of detectors: for example, SCUBA on the
JCMT~\citep{Holland99}, MAMBO on the IRAM 30~m
telescope~\citep{Kreysa98}, and Bolocam on the Caltech Submillimeter
Observatory~\citep{Haig2004,Glenn2003}.  AzTEC is a
millimetre-wavelength bolometer camera designed for the Large
Millimeter Telescope (LMT). Its 144 silicon nitride micromesh
bolometers operate with a single bandpass centered at either 1.1, 1.4,
or 2.1~mm, with one bandpass available per observing run. AzTEC was
commissioned at 1.1~mm during an engineering run at the James Clerk
Maxwell Telescope (JCMT) in June 2005 and completed a successful
observing run at the JCMT from November 2005-February 2006 during the
JCMT05B semester.  In 2007 the instrument was installed on the 10~m
Atacama Submillimeter Telescope Experiment (ASTE) where it will reside
as a facility instrument through 2008.  AzTEC will be installed on the
LMT in early 2009.

In its 1.1~mm wavelength configuration, AzTEC is sensitive to the
Rayleigh-Jeans tail of the thermal continuum emission from cold dust
grains. Consequently, AzTEC is well-suited for extragalactic surveys
of dusty, optically obscured starburst (or AGN-host) galaxies detected
by their sub-mm/mm emission (so-called Sub-Millimetre Galaxies, or
SMGs).  As the far-infrared (FIR) peak of the cold dust emission is
increasingly redshifted to mm-wavelengths with increasing distance,
AzTEC is equally sensitive to dusty galaxies with redshift $z\ge 1$.
The AzTEC/JCMT05B surveys of the Lockman Hole and the Subaru Deep
Field~(Austermann {\it et al.} in preparation),
COSMOS~\citep{scott2008}, and GOODS-North~(Perera {\it et al.} in
preparation) represent the largest SMG surveys with uniform high
sensitivity ($1\sigma \sim 1$ mJy) ever conducted, successfully
demonstrating the superb mapping speed and stability of the instrument
and providing a new insight on the clustering and cosmic variance of
the SMG population.  In a similar vein, AzTEC's high mapping speed and
high sensitivity make it an interesting instrument for studies of cold
dust in the Milky Way and in nearby galaxies.

Configured for 1.4~mm and 2.1~mm observations, AzTEC will be tuned to
make high resolution images of the Sunyaev-Zel'dovich effect in
clusters of galaxies.  AzTEC on the LMT will have a per-pixel
resolution of 10$^{\prime\prime}$ at 2.1~mm (5$^{\prime\prime}$ at
1.1~mm) and will therefore be an unprecedented instrument for the
study of the energetics of the free electron gas in clusters.

In this paper we report on the design and performance of the AzTEC
instrument. In Section~\ref{sec:system}, we describe the components of
the instrument and its configuration. The observing software, mapping
strategies, and practical observing overheads are addressed in
Section~\ref{sec:obs}. We describe the calibration of AzTEC data in
Section~\ref{sec:cal} and in Section~\ref{sec:per} we list the
sensitivity and mapping speeds of AzTEC as measured at the during the
JCMT05B run. We conclude with a brief discussion of the future of
AzTEC in Section~\ref{sec:fut}.

\section{SYSTEM DESIGN}
\label{sec:system}
\begin{figure}
\includegraphics[width=\hsize]{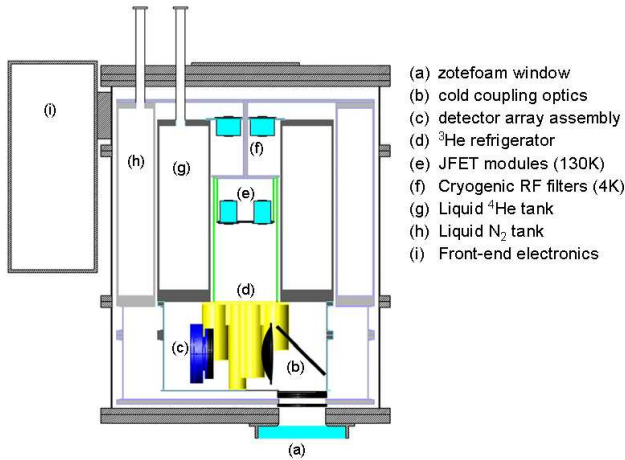}
\caption{A cut-away view of the AzTEC cryostat.  Mechanical supports have been removed for clarity.}
\label{fig:overview} 
\end{figure}

A cut-away view of the AzTEC instrument is shown in
Figure~\ref{fig:overview}.  Each subsystem labeled in the
figure is described in detail below.

\subsection{Cryostat and Cryogenics}
\label{sec:cry}

The AzTEC detector array is cooled using a three-stage, closed-cycle
$^{3}$He refrigerator~\citep{Bhatia2000} mounted to a liquid cooled
4~K cryostat.  Operation of the AzTEC refrigerator is fully automated
and remotely controlled.  The refrigerator takes approximately 140
minutes to cycle.  Optimization of the fridge cycle has led to stable hold
times well over 24 hours at sea level with a full optical load.  The
operational hold time of the refrigerator was 36 hours at the JCMT
(4092~m) and 42 hours at
the higher elevation ASTE telescope (4860~m).

The refrigerator's ultra-cold (UC) stage, operating at a temperature
of approximately 250~mK ($256.5\pm1.9\pm10$~mK where the 10~mK
uncertainty comes from uncertainty in the calibration of the GRT),
provides the thermal sink to the bolometer array.  An intercooler (IC)
stage, operating at approximately 360~mK, heat sinks an intermediate
temperature stage of the detector array support structure to thermally
intercept heat flowing along the mechanical and electrical connections
to the UC stage.  Both the IC stage and the detector array assembly
are thermally isolated and mechanically supported using two sets of
short, hollow Vespel standoffs in a radial configuration.  The
resulting structure is extremely compact and stiff and the symmetry of
the system prevents net motion of the optical axis under thermal
contraction.  The weakest vibrational mode of the support is along the
optical axis and has a resonant frequency $>400$~Hz.  The UC and
IC temperatures showed no correlation with cryostat tilt or effective
atmosphere optical depth during the JCMT observations.

The custom designed AzTEC cryostat is a 'wet' system with liquid
helium (23 liters) and liquid nitrogen (26 liters) tanks.  The design
and preparation of the cryostat resulted in operational (full loading,
including daily fridge cycles) hold times of 3 days for liquid helium
and over 7 days for liquid nitrogen.  The cryogen tanks are concentric
annuli with near optimum length/diameter ratios to maximize their hold
times.  The inner diameter of the helium tank is 15.25~cm -- a feature
that adds valuable volume and accessibility to the shielded 4 K
workspace without compromising cryogen capacity.

One challenge of working with semiconductor thermistors is the need to
impedance transform using warm JFETs.  AzTEC's low-noise U401 JFETs
amplifier are suspended inside the inner diameter of the liquid helium
tank by a set of reentrant G-10 tubes with tuned conductivity to allow
the stage to self-heat to 130~K where the JFETs have empirically been
found to exhibit low voltage and current noise.  An intermediate stage
of the suspension is sunk to the 77~K bath and intercepts the heat
dissipated by the JFETs.  These 130~K and 77~K components are
radiatively insulated with several layers of multi-layer insulation
(MLI).  The inner wall of the 4~K helium tank is painted with an
IR-black paint to absorb stray thermal radiation.

The cryostat is designed to minimize EMI/RFI susceptibility.  Two
nested high conductivity faraday shields are formed through which all
electrical connections pass via imbedded pi-filters.  The ``cleanest''
volume is bounded by the inner shield which surrounds the 4~K work
volume.  Custom compact in-line cryogenic pi-filters were constructed for AzTEC
in coordination with the manufacturer.  The outer faraday cage is formed by the
cryostat vacuum jacket and electronics enclosure.

\subsection{Detectors}
\label{sec:det} 
The AzTEC array is a 76~mm diameter monolithic silicon wafer
containing 151 silicon nitride micro-mesh (spider-web) bolometers with
neutron transmutation doped (NTD) Ge
thermistors~\citep{Bock1996,Mauskopf1997,Turner2001}.  The wafer
is organized into six pie shaped regions (hextants) that together
contain the 144 optically active bolometers and their wiring.  The
detectors are arranged in a close-packed hexagonal configuration with
a spacing set by the detector feed-horn aperture of 5~mm.  The bias
for each detector is provided via a symmetric pair of 10~M$\Omega$
resistors contained in a separate module located close to the detector
array.  The 7 ``blind'' bolometers of the array are not biased or
read out.

The bolometer behavior is dictated by three sets of parameters: the
thermistor properties, the amount of absorbed optical and electrical
power, and the thermal link between the bolometer and the cold heat
sink of the cryostat.  The thermal link is defined by a Au film
deposited on one of the mesh support legs connected to the wafer
substrate.  Its heat conductance has the form $g_0 T^\beta$ with
$\beta$ in the range 1.2-1.6~\citep{Haig2004,Glenn2003} where $T$ is
the bolometer operating temperature.  The thermistor resistance has
the form $R_0
\mathrm{exp} \sqrt{\Delta/T}$ with targeted $R_0$ and $\Delta$ values
of 100~$\Omega$ and 42~K respectively.  In principle, $g_0$ as well as
the electrical bias power are chosen to minimize detector noise given
a known optical loading.  For the AzTEC array we chose a
conservatively high $g_0$ that results in a heat conductance of
167~pW/K at 300~mK to mitigate the effects of any unanticipated excess
optical loading and as a compromise for the different amounts of
loading expected in the three possible optical passbands.  For
this choice, the noise equivalent power (NEP) predicted from our
bolometer model for the optically loaded detectors is $5-7\times
10^{-17}$ W/$\sqrt{\rm{Hz}}$, which is comparable but sub-dominant to
the photon-noise background limit (BLIP) in the 1.1~mm passband for
anticipated operating conditions.  The expected detector time
constant, $\tau$, of 3-4~ms also applies over the range of reasonable
operating conditions.  The thermistor resistance and bias resistor
values are chosen such that the detector and photon noise are dominant
over noise sources from circuit elements further down the readout
chain.

Once AzTEC is installed on a telescope, the only free parameter for
optimizing sensitivity is the amount of electrical (bias) power
dissipated at the bolometer and hence the bolometer operating
temperature $T$.  Following \citet{mather84a}, the balance between
phonon noise, which rises with $T$, and Johnson noise, which falls
steeply with rising $T$, results in an optimum value of the bias
voltage for a given set of detector properties and a given optical
loading.  In practice since there is a small spread in detector
properties across the array and a slowly varying, but unpredictable,
optical loading due to the atmosphere, we fixed the thermistor bias
amplitude for all detectors in all hextants at 62.5~mV over the entire
JCMT05B observing run.  For an atmospheric optical depth at 225~GHz,
$\tau_{225}$, of 0.1, this conservatively high bias results in a
sensitivity 10\% worse than expected for a bolometer that has the
design parameters.  As shown in Section~\ref{sec:per}, nightly load
curves, noise estimates, and beam map observations during the JCMT05B
run indicate that the detector sensitivity was near optimum over the
entire observing run for this choice of bias.

\subsection{Optics}
\label{sec:opt} 

\begin{figure*}
\includegraphics[width=\hsize]{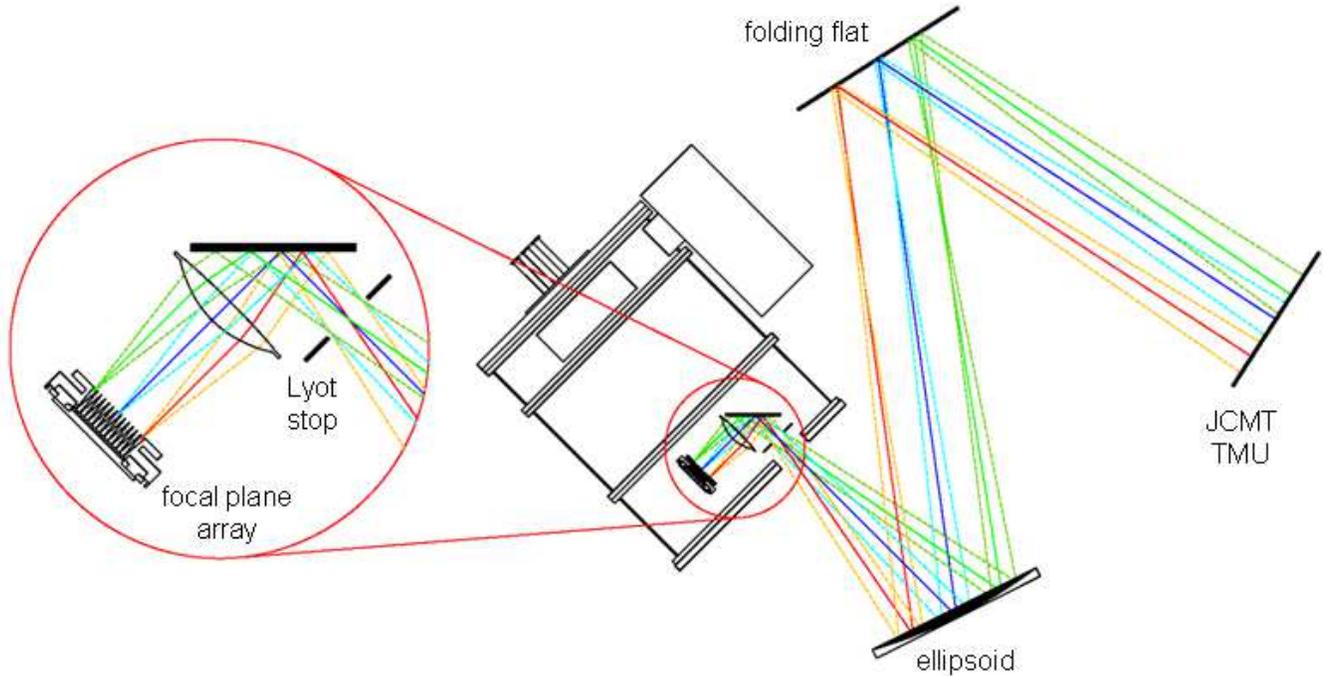}
\caption{Schematic of AzTEC mounted in the JCMT receiver cabin with the 
         associated coupling optics.  All filters and mechanical
         supports have been omitted for clarity.  Zemax produced rays
         are shown for the central and outermost detectors in the
         plane of the drawing to illustrate the optical path of the
         various beams.  The Cassegrain focus of the JCMT is located
         at the convergence of the beams just below the folding flat.}
\label{fig:jcmt} 
\end{figure*}

A schematic diagram showing the relative positions of the optical
elements with AzTEC in the JCMT receiver cabin is shown in
Figure~\ref{fig:jcmt}.  Our description of the coupling optics will
start from the JCMT's tertiary mirror unit (TMU) located inside the
receiver cabin.  A complete listing of JCMT optical specifications may
be found on the JCMT web site.  A list of optical constants for the
AzTEC instrument and its JCMT coupling optics is given in
Table~\ref{tab:optical_design}.

At the JCMT, photons from the secondary mirror reflect off of the TMU
which is located behind the vertex of the primary mirror.  For the
AzTEC system the TMU is oriented to direct the incoming beam upwards
at an angle of $53.2^\circ$ with respect to the vertical.  A flat
folding mirror then directs the beam downwards to an ellipsoidal
mirror which converts the $f/12$ beam to an $f/11.8$ beam and directs
the beam into the cryostat.  The JCMT's cassegrain focus resides
between the folding flat and the ellipsoidal mirror.  The AzTEC dewar
is located such that a blackened Lyot stop, at a physical temperature
of 4~K, is coincident with an image of the primary mirror.  The image
of the primary has a diameter of 52~mm and so the Lyot stop, which has
a diameter of 50~mm, provides a cold guard ring to minimize the
spillover at the primary mirror edge while allowing an aggressive edge
taper.  The folding flat and ellipsoidal coupling mirror are sized
such that the edge taper of the outermost beam on each surface is less
than -30dB.  Both mirrors and the cryostat are supported rigidly to
minimize optical microphonic pickup.

Inside the cryostat, the detector array assembly is situated behind a
4~K biconvex aspheric lens fashioned from ultra-high molecular weight
polyethylene (UHMWPE) with index of refraction, $n=1.52$.  The lens
surface is grooved with a series of concentric trenches of depth
$\lambda/4\sqrt{n}$ which results in an anti-reflection surface tuned
for the system passband.  Considering the optical path in reverse,
chief rays from each of the the detectors' feedhorns are focused by
the lens onto the center of the blackened 4~K Lyot stop so that all
detectors illuminate the 4~K edge of the Lyot stop with an edge taper
of $-5.2$~dB in power.  A 4~K folding mirror between the lens and the
Lyot stop allows for a long optical path in a compact configuration to
aid in keeping the overall structure mechanically rigid and
isothermal.

The 250~mK bolometer array is mechanically and thermally supported
between monolithic gold-plated aluminum arrays of tuned backshorts and
feedhorns in a manner identical to that of the Bolocam
instrument~\citep{Glenn2003}.  The array of backshorts and integrating
cavities sits $\lambda/4$ behind the detectors in order to optimize
detector absorption and minimize optical cross-talk between
detectors~\citep{Glenn2002}.  The array of single-moded conical feed
horns lies in front of the detectors and produces a set of
quasi-gaussian beams, each with gaussian beam radius, $w_a$, of
1.84~mm at the horn aperture for the 1.1mm wavelength configuration.
This corresponds to an $f/$\# for each horn of $f/3.2$.

\begin{table}
\begin{center}
\begin{tabular}{|l|c|}
\hline
ellipsoid focal length           & 645.2~mm  \\
ellipsoid reflection angle       & 37$^\circ$\\
UHMWPE lens focal length         & 163.6~mm  \\
$f/$\#$_{\mathrm{cass}}$           & 12.0      \\
$f/$\#$_{\mathrm{Lyot}}$           & 11.8      \\
$f/$\#$_{\mathrm{horns}}$          & 3.2       \\
Lyot stop diameter               & 50~mm     \\
edge taper at Lyot stop          & -5.2dB    \\
image of primary mirror dia.     & 52~mm     \\
plane wave coupling efficiency at Lyot stop  & 0.62 \\
detector spacing                 & 1.4$f\lambda$ \\
\hline
\end{tabular}
\vspace{-.25cm}
\end{center}
\caption{AzTEC design optical parameters for coupling to the JCMT.  
         The calculation of the plane wave coupling efficiency at the
         Lyot stop is made assuming gaussian optics and following the
         perscription of \citep{Goldsmith1998}.}
\label{tab:optical_design}
\end{table}

\subsubsection{Optical Filtering}

\begin{figure}
\includegraphics[width=\hsize]{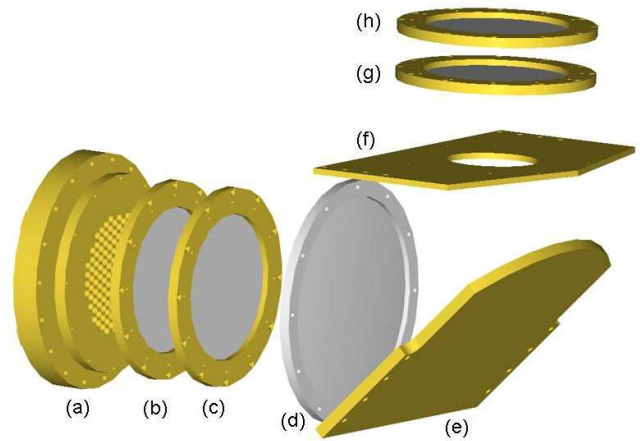}
\caption{An exploded layout of the internal AzTEC optics.  Components are: 
         (a) integrated detector array, backshorts, and conical
         feedhorns (250~mK); (b) 310~GHz low-pass filter (250~mK); (c)
         360~GHz low-pass filter (250~mK); (d) bi-convex ultra-high
         molecular weight polyethylene lens (4~K); (e) folding flat
         (4~K); (f) 50~mm diameter Lyot stop (4~K); (g-h) 390~GHz and
         1050~GHz low-pass filters (4~K).  Not shown are an additional
         77~K 540~GHz low-pass filter and a Zotefoam PPA-30 cryostat
         window.  }
\label{fig:optics} 
\end{figure}

An exploded view of the internal optical layout including the filters is
shown in Figure~\ref{fig:optics}.  A short stub of single-mode
waveguide (3.4~mm length for the 1.1~mm passband configuration) at the
detector end of each horn provides the high-pass filter to the system
passband.  All of the low-pass filters in the system are quasi-optical
metal mesh filters consisting of a series of resonant metallic meshes
separated by transmission line sections. The resonant mesh filters are
fabricated from thin copper films deposited on plastic (polypropylene
or mylar) substrates and patterned into inductive or capacitive grids
\citep{Tucker2006,Ade2006}.  In the 1.1~mm configuration, a set of
low-pass filters, one rolling off at 360~GHz and the other rolling off
at 310~GHz, are mounted to the feedhorn array and define the upper
portion of the passband.  For configurations with other band centers,
the backshort array, the feedhorn array, and these two low-pass
filters are replaced.

\begin{figure}
\includegraphics[width=\hsize]{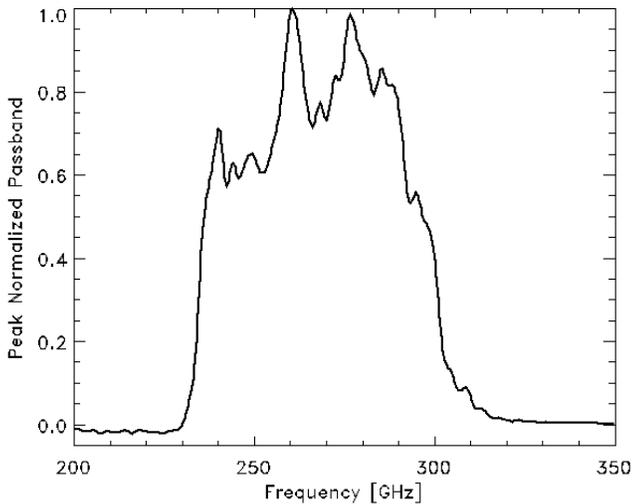}
\caption{The AzTEC system bandpass for a flat spectrum input source, normalized to the peak response.  
         (The low frequency portion of the spectrum is slightly
         negative due to a small phase offset in the fourier transform
         spectrometer used to measure the response.)  }
\label{fig:bandpass} 
\end{figure}

Further optical filtering is required to suppress harmonic response
from the 250~mK low-pass filters.  A low-pass capacitive mesh filter
(390~GHz edge) and an anti-reflection coated 1050~GHz low-pass filter
are mounted on the telescope side of the Lyot stop at 4~K.  A 77~K,
540~GHz low-pass filter is mounted to the liquid nitrogen shield 2~cm
further down the optical axis.  Finally, a room temperature Zotefoam
PPA30 (Zotefoams PLC) window provides IR scattering and a vacuum seal
that is transparent at 1.1~mm wavelengths to better than 1\% according
to in-lab measurements with a Fourier Transform Spectrometer.  The
resulting system passband is shown in Figure~\ref{fig:bandpass}.

Measured optical characteristics are described further in
Section~\ref{sec:performance}.

\subsection{Signal Chain}
\label{sec:ele} 

\begin{figure*}
\begin{center}
\includegraphics[width=16cm]{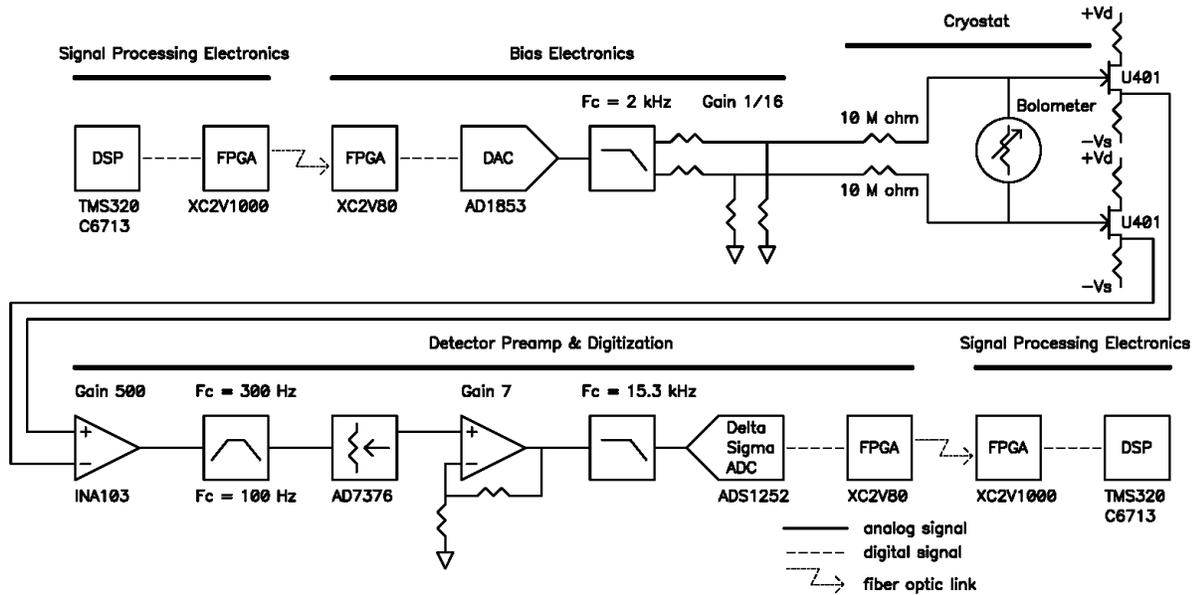}
\caption{Schematic of the detector readout chain. All analog electronics are 
         replicated 144 times in the sysem.  }
\label{fig:elec}
\end{center} 
\end{figure*}

A schematic for the detector signal chain is shown in
Figure~\ref{fig:elec}.  Each of the 144 bolometers is pseudo-current
biased by a differential 200~Hz sine wave produced by a digital to
analog converter (DAC) -- one DAC per hextant -- and a pair of
10M$\Omega$ load resistors.  Noise in the bias is correlated across
all detectors of a hextant and is removed in the data analysis.  The
bias frequency of 200~Hz is chosen to be fast compared to the detector
time constant, $\tau>3$~ms, but slow enough to avoid excessive
attenuation from the effective low-pass RC filter arising from the
bolometer/load resistor combination ($\sim 3$~M$\Omega$) and the net
parasitic capacitance of the signal lines and JFET gates
($C_\mathrm{parasitic}\sim 60-110$~pF).

Following each bolometer is a matched pair of low-noise U401 JFETs, in
the source-follower configuration, that transforms the detector output
impedance from several M$\Omega$ to $\sim 350~\Omega$.  Signals from
the JFET amplifiers exit the cryostat via pi-filters at 4~K and 300~K
and are read out by room-temperature, low-noise (1.7~nV~Hz$^{-1/2}$)
instrumentation amplifiers (Texas Instruments, type INA103) with
a fixed gain of 500.  Each detector signal is then high-pass filtered
at 100Hz and low-pass filtered at 300Hz before a programmable
attenuation is applied by a digitally controlled resistive divider
(see Figure~\ref{fig:elec} for part numbers.)  The output of the
divider is then digitized at 6.144~MHz by high-precision 24-bit
delta-sigma A/D converters (Texas Instruments ADS1252) which produce
an output data stream at a sampling rate of 16 kHz.

Once digitized, a Field Programmable Gate Array (FPGA) serializes each
hextant's data into a single data stream.  All
processing up to this point happens in a custom electronics enclosure
attached to the cryostat (the ``front-end'' electronics).  This data is
sent via fiber optics to the off-cryostat back-end electronics
where a single FPGA collects all six hextants' data streams.  The back-end FPGA
presents this data in parallel to a single Digital Signal
Processor (DSP) for demodulation of each channel (twice) using the
reference bias signal and its quadrature phase.  The DSP caches the
demodulated data streams in memory and applies a sharp block
Finite Impulse Response filter, whose cutoff is just below the
Nyquist frequency of the 64~Hz final sampling rate.

Bias generation and housekeeping electronics are also part of the
front-end electronics located at the cryostat.  The bolometer bias
signal is generated from a digital sine wave stored in memory in the
back-end FPGA and continuously fed via the fiber-optic connection to
the DACs that generate the analog sine wave for each hextant.  The
bias amplitude and phase are controlled digitally and in a real-time
fashion. The bias frequency for the JCMT05B
observations was 200~Hz.  The Housekeeping card reads out
internal thermometer voltages, provides power, and controls the $^3$He
refrigerator.

The back-end electronics are controlled by a dedicated Motorola
Power-PC running the real-time VxWorks operating system.  While a
real-time system is not required by the data acquisition architecture, a
real-time system simplifies synchronization of the native AzTEC
signals with telescope pointing and other environmental signals.  From
the user's perspective, AzTEC is controlled and data is stored using a
standard Linux PC running the Large Millimeter Telescope Monitor and
Control (LMTMC) system which was developed at UMass-Amherst for
general use by all LMT instruments~\citep{souccar2004}.  All
communication and data transfer between the Linux machine and the
Power-PC is via dedicated Ethernet.  Including pointing signals and
telescope housekeeping signals, the total data rate of AzTEC is $\sim
80$~kB/s.

\section{OBSERVING}
\label{sec:obs} 
The AzTEC data set is logically grouped into ``observations'' which
include engineering tests (e.g., calibration, pointing, or focus
observations) as well as scientific observations.  All data is stored
in a machine independent binary format, Network Common Data Form
(NetCDF).
AzTEC NetCDF files are self-describing in that they include all
information required to produce an optimal image, including header
information defining the data as well as all observing parameters.

Real-time monitoring of AzTEC is conducted using various tools in the
LMTMC package and the KST plotting
suite.  Quick-look software, custom written in the IDL programming
language, is used to produce images immediately after the completion
of each observation.  These simple maps allow the user to quickly
evaluate telescope pointing and focus, and provides an assessment of
the overall quality of the data.

\subsection{Observing Modes}
\label{obs:mod}

AzTEC is a passive instrument with respect to the telescope and
operates completely independently of the observing mode.  Since the
mapping efficiency is a strong function of the observing mode and
observing parameters for maps on-order the field of view of the
instrument, we have adopted two primary modes in the existing data
analysis suite.

\subsubsection{Jiggle-Mapping}
\label{mod:jig}

In the jiggle mapping mode, the secondary mirror is chopped between
on-source and off-source sky positions while stepping through a series
of small offsets (jiggles) to fill in gaps in coverage left by the
approximately 20\arcsec spacing of the detectors.
Differencing chopped data at a single jiggle position removes the
effects of low-frequency atmospheric and instrumental drifts.  Chop
frequencies during the JCMT05B run ranged from 2 to 4~Hz with a
secondary mirror transit time of less than 30~ms, and a jiggle
frequency of 1~Hz.  Once the full jiggle pattern is completed, the
primary mirror ``nods'' to put the off-source beams on source.
Averaging data from opposite nods further suppresses differential
pickup from spillover at the primary mirror, temperature gradients on
the primary, and scattering from the secondary support structure.

\begin{figure}
\includegraphics[width=\hsize]{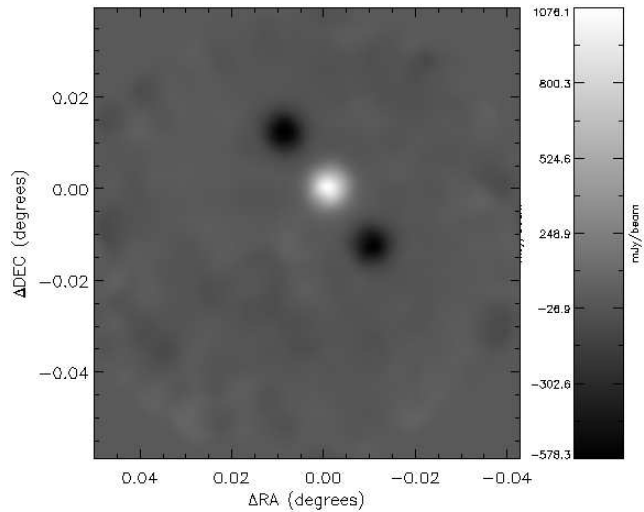}
\caption{\label{fig:jig} Jiggle-map of QSO J1048+7143. 
A gaussian fitted to the peak flux gives the telescope boresight
  offset from the reference bolometer (Section \ref{ovr:pnt}).}
\end{figure}

\begin{figure}
\includegraphics[width=\hsize]{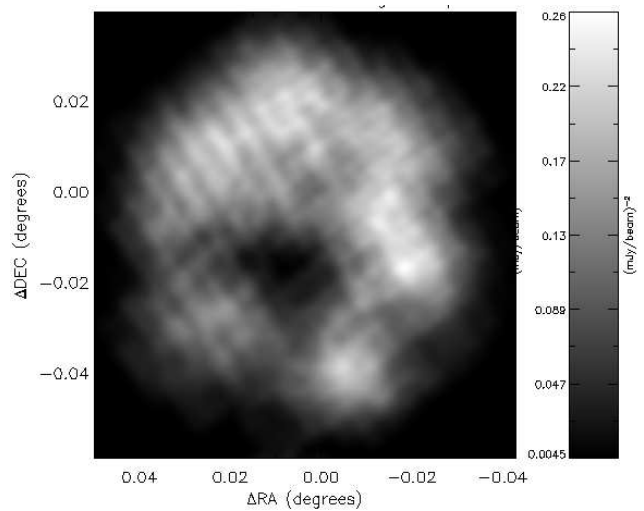}
\caption{\label{fig:jigweight} The corresponding weight map 
(in mJy/beam$^{-2}$) for the signal map in Figure~\ref{fig:jig}.  The
lack of coverage due to a clustering of inactive detectors is indicated
by the low weight region near the center of the map.}
\end{figure}

For fields on the same scale as the array field of view (FOV) or
smaller, jiggle-mapping is the highest efficiency observing mode for a
given target sensitivity.  However, when imaging fields that are much
larger than the array FOV, the overheads of jiggle mapping and the
high variability in coverage make scan mapping essential.  The
coverage map from a jiggle-map is highly sensitive to the distribution
of detector sensitivities on the array. This is particularly
evident in AzTEC jiggle-maps taken during the JCMT05B observing run, as
shown in the Figures~\ref{fig:jig} and~\ref{fig:jigweight}, where a
contiguous region of detectors were inoperable.  This resulted in a
portion of the field being severely undersampled.  For jiggle-map
observations of sources with a known location, we adjust for the
non-uniform coverage by offsetting the boresight pointing of the
telescope so that the target falls in a region fully populated by
working detectors.

\subsubsection{Raster-Scanning}
\label{mod:ras}
For raster-scan observations the telescope boresight sweeps across the
sky along one direction, takes a small step in the orthogonal
direction, and then sweeps back along the next row of the scan. This
pattern is repeated until the entire field has been imaged. Because of
the low-frequency stability of the detectors, we do not chop the
secondary mirror. In fields where no bright sources are expected,
removal of the atmospheric contamination is currently accomplished via
a principal component analysis (PCA) of the full data stream on a scan
by scan basis~\citep{Laurent2005}.  The PCA technique is used in the
analyses leading to the instrument performance and characteristics
detailed in Section~\ref{sec:per}.

\begin{figure}
\includegraphics[width=\hsize]{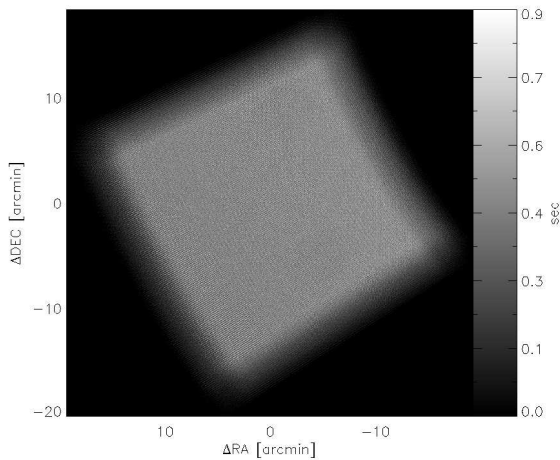}
\caption{\label{fig:ras} The integration time in each 2\arcsec pixel for a
 25$^{\prime}\times$25$^{\prime}$ raster-scan map with scanning done
  in the elevation direction and with 10\arcsec step sizes in azimuth,
  demonstrating the uniform coverage achieved
  with this observing mode. }
\end{figure}

The major benefit to raster-scanning is in the ability to map a large
area of sky in a single observation with very uniform coverage, as
shown in Figure \ref{fig:ras}. The distribution of inoperable
detectors on the array only affects the ultimate sensitivity of the
map, and not the uniformity of the coverage in the map.  Scan speeds
are limited by the detector time constant and stability of the
telescope but are generally chosen based on the science target and the
atmospheric opacity.  For point source observations at the JCMT05B
run, high scan speeds were preferred in order to move the signal
bandwidth above the knee frequency of residual atmospheric
contamination (see Figure~\ref{fig:sens}), however at a cost in
overall observing efficiency due to the fixed length turnaround time
(5~s) of the telescope.  Coadding tens of raster scanned maps taken
with different sky position angles reduces scan-synchronous effects in
the maps and offers excellent cross-linking of the data.

\subsection{Observing Overheads}
\label{obs:ovr}

While AzTEC was on the JCMT, the following ancillary observations were
made, averaging 24.7\% of the total available observing time.

\subsubsection{Focus Observations}
\label{ovr:foc}
Focus measurements consist of a series of jiggle-map observations on a
bright ($\sim$few Jy) point source as the secondary mirror steps
through different focus settings. For each focus setting, we fit a
2-dimensional Gaussian to the region of the image containing the
source.  The optimal focus location is the secondary position where
the beam full-width at half maximum (FWHM) is minimized and the peak
amplitude of the signal is maximized.  For the JCMT05B run 
we focused the telescope at the beginning and mid-way through each
night of observing.

\subsubsection{Relative Bolometer Offsets}
\label{ovr:bmp}

Since the array orientation is fixed in azimuth and elevation, the
relative offset on the sky between any two detectors is constant. We
determine these offsets by mapping a bright point source each evening
prior to science observations and after focusing the telescope.  A
high-resolution map is made such that the point spread function (PSF)
of each detector in the array is sampled with at least
4$^{\prime\prime}$ resolution in order to determine the relative
bolometer positions to an accuracy of $\approx 5\%$ of the PSF FWHM.
These ``beam map'' observations also provide the absolute calibration of detectors
in the array (see Section~\ref{sec:cal}).

\subsubsection{Loadcurves}
\label{ovr:ldc}
Loadcurve measurements -- sweeping the detector bias through its full
range of commandable values while viewing a blank patch of sky -- are
made each evening following the determination of the relative
bolometer offsets.  From the load curves we determine the total
optical power, $Q$, absorbed by each detector as well as the
responsivity, $S$, (the conversion from Volts read out of the detector
to Watts absorbed) of each detector.  By measuring the responsivity in
this manner under a wide range of atmospheric opacities we construct a
correction to the non-linearity of the detector response due to the
overall variation in the atmospheric optical loading.  This process is
described further in Section~\ref{cal:res}.

\begin{figure}
\includegraphics[width=\hsize]{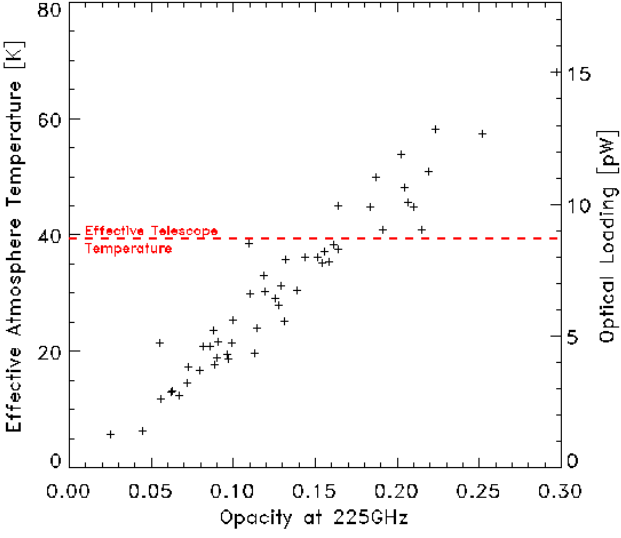}
\caption{\label{fig:tau_v_atm} The effective atmosphere temperature in 
the 1.1~mm AzTEC bandpass as a function of the atmospheric opacity at
225~GHz as measured by the CSO.  The axis on the right gives the corresponding amount of
optical power absorbed by the AzTEC detectors.  The horizontal line
shows the effective temperature/power of the telescope and coupling
optics.}
\end{figure}

Figure~\ref{fig:tau_v_atm} shows the contributions of the atmosphere
and other sources to the total optical power absorbed by the AzTEC
detectors.  The telescope, coupling optics, and any parasitic optical
loadings on the detectors deliver a combined 9~pW of power --
corresponding to an effective blackbody temperature of 39~K.  The
optical loading due to the atmosphere is linear with the
opacity as measured by the Caltech Submillimeter Observatory (CSO) tau
meter at 225~GHz over the range of opacities suitable for observing.

\subsubsection{Pointing Observations}
\label{ovr:pnt}
Pointing measurements are performed approximately every hour in order
to measure the absolute pointing offset between the telescope
boresight and a reference bolometer. An optimal pointing source is
bright ($\ge1$~Jy), unresolved, and located near the science
target. Pointing observations typically bracketed a series of science
observations so that slow drifts in the residuals to the telescope pointing
model could be fit out.  Since only a few bolometers must pass over
the source, pointing observations are usually carried out in
jiggle-map mode. We fit a 2-dimensional Gaussian to the point source
image, and the best-fit location of the peak signal gives the
boresight offset as shown in Figure \ref{fig:jig}. To correct the
pointing signals for a given science observation, a pointing model is
generated by interpolating between the pointing measurements taken
over a night.

\section{CALIBRATION}
\label{sec:cal} 
The output of bolometer $i$ at sky position $\alpha$, 
$b_{i,\alpha}$ (in units of V), is given by

\begin{equation}
\label{eqn:bs1}
b_{i,\alpha} = 
S_i(Q) A_{\rm{eff}}
\eta \int_{0}^{\infty} d\nu f(\nu) 
\int_{\rm{sky}} d\Omega P_i(\Omega_{\alpha}-\Omega)
e^{-\tau_{\rm{eff}}} I_{\nu}(\Omega),
\end{equation}
where $S_i(Q)$ is the responsivity (in V/W), $Q$ is the optical
loading (in W) dominated by the telescope and atmosphere,
$A_{\rm{eff}}$ is the effective telescope aperture, $\eta$ is the
optical efficiency, $f(\nu)$ is the peak-normalized AzTEC bandpass,
$P_i(\Omega_{\alpha}-\Omega)$ is the peak-normalized AzTEC beam
pattern for bolometer $i$ at sky position $\Omega_{\alpha}$,
$\tau_{\rm{eff}}$ is the opacity, and $I_{\nu}(\Omega)$ is the source
intensity on the sky (in Jy beam$^{-1}$). As discussed below, $S_i$
and $\tau_{\rm{eff}}$, both of which depend on observational
conditions (e.g. weather) and change significantly on the time scale
of hours, are modeled as functions of the ``dc'' level of the
bolometer signal.

The flux conversion factor for bolometer $i$, FCF$_i$ (in units of Jy
beam$^{-1}$W$^{-1}$), is an expression involving all factors that are,
in principle, constant (i.e. source and weather independent) over the
entire observing run, and is defined as

\begin{equation}
\label{eqn:fcf}
\mbox{FCF}_i = {1 \over A_{\rm{eff}} \eta \int_{0}^{\infty}
  d\nu \hspace{0.1cm}f(\nu) \int_{sky} d\Omega
  \hspace{0.1cm}\mbox{P}_i(\Omega_{\alpha}-\Omega)}.
\end{equation}
For a point source located at ($\theta_0$,$\phi_0$) with average flux
density $\bar{I}$ over the optical bandpass, the bolometer output is

\begin{equation}
\label{eqn:bs2}
b_i(\theta_0, \phi_0) = {S_i(Q)
e^{-\tau_{\rm{eff}}} \bar{I} \over \mbox{FCF}_i}.
\end{equation}

\subsection{Responsivity and Extinction Corrections}
\label{cal:res}
\begin{figure}
\includegraphics[width=\hsize]{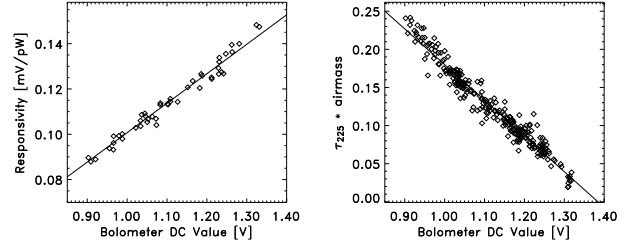}
\caption{\label{fig:rat} Left: Responsivity versus bolometer dc-level 
  for a typical detector as determined from all of the loadcurves
  taken during the JCMT05B run.  A best-fit line is
  over-plotted. Right: Opacity versus bolometer dc-level for the same
  detector.}
\end{figure}

The optical loading on the detectors and the bolometer responsivities
are determined from loadcurve measurements (see Section
\ref{ovr:ldc}). For the range of total power loading on the detectors
observed during the JCMT05B run, the responsivity is linearly
proportional to the demodulated dc-level of each bolometer's
timestream signal. The responsivity of a typical bolometer versus the
dc-level measured from all of the loadcurve observations taken during the
JCMT05B run is shown in the left panel of Figure \ref{fig:rat}. The solid
curve shows the linear fit to the measurements. We derive the best-fit
offset and slope for each bolometer separately since the spread in
these parameters is large compared to the formal errors on the fits.

The atmospheric extinction $e^{-\tau_{\rm{eff}}}$ is corrected in a
similar way.  A linear correlation exists between the atmospheric
opacity, $\tau_{\rm{eff}}$ and the bolometer dc-signals (see right
panel of Figure~\ref{fig:rat}).  For the JCMT data we use the
atmospheric opacity at 225~GHz as determined from the CSO tau monitor,
which records $\tau_\mathrm{225 GHz}$ (zenith) every 10 minutes, to
calibrate the relation with the bolometer dc-levels.

\subsection{Measured FCF from Beam Map Observations}
\label{cal:bmp}
To determine the flux conversion factor for each bolometer, FCF$_i$,
we beam map a primary or secondary flux calibrator each night. The
timestream bolometer signals are corrected for extinction and the
responsivity is factored out.  An iterative cleaning technique is used
to minimize errors in fitted parameters due to atmospheric
contamination.  In the final iteration, maps are made for each
bolometer separately and we fit a 2-d Gaussian to each map. The
best-fit amplitude combined with the known flux of the source gives
the FCF for each detector.

To correct the flux for the angular size of Uranus, we assume that
Uranus is a disk with angular radius $\Theta_{\rm{U}}$ and brightness
temperature T$_b$ so that $I(\Omega)$ = T$_b\Phi(\Omega)$, where
$\Phi(\Omega)=1$ for $\theta\le \Theta_{\rm{U}}$ and 0 otherwise and
T$_b=92.6\pm1.7$~K~\citep{gri93}. The average flux of Uranus for
bolometer $i$ is then given by

\begin{equation}
\label{equ:ufl}
\bar{I} = \frac{\int_{0}^{\infty} d\nu \hspace{0.1cm}f(\nu) {2kT_{b,o} \nu^2 \over c^2} 
\int_{sky} d\Omega \hspace{0.1cm} P_i(\Omega_{\alpha}-\Omega) \Phi(\Omega)}
{\int_{0}^{\infty}
  d\nu \hspace{0.1cm}f(\nu) \int_{sky} d\Omega
  \hspace{0.1cm}\mbox{P}_i(\Omega_{\alpha}-\Omega)},
\end{equation}
which can then be used in Equation~\ref{eqn:bs2}. Uranus is small
compared to the AzTEC detector PSFs on the JCMT ($2\Theta_{\rm{U}}\ll
\theta_{b_i}$, where $\theta_{b_i}$ is the true beam FWHM for
bolometer $i$) and so will appear approximately Gaussian with a
measured FWHM $\theta_{m_i}$, where

\begin{equation}
\label{equ:bea}
\theta_{b_i}^2 = \theta_{m_i}^2 - {\ln{2} \over 2}(2\Theta_{\rm{U}})^2
\end{equation}
\citep*{baa73}. We measure $\theta_{m_i}$ from Uranus beam maps, and
use Equation \ref{equ:bea} to determine $\theta_{b_i}$ and
subsequently P$_i$($\Omega$). We then calculate the integral in Equation
\ref{equ:ufl} to determine the correction factor for the flux of
Uranus.

For beam maps of Uranus, there is a statistically significant increase
in the measured FCF$_i$ for measurements taken within one hour after
sunset at the JCMT.  Measurements taken after this time have
constant FCF$_i$. This is consistent with rough estimates of the
telescope's thermal time constant. For this reason, we determine the
average flux conversion factor for each bolometer $<$FCF$_i$$>$ by
averaging over all FCF$_i$ measured from Uranus beam maps taken $\ge$1
hour after sunset, and we use these values to calibrate all science
observations taken after the telescope has settled.  A linear
correction factor derived from beam maps taken within one hour after
sunset is applied to science data taken during this period.  We model
this correction factor f $\equiv$ FCF$_i$/$<$FCF$_i$$>$ as a linear
function of the time after sunset, such that

\begin{eqnarray}
\label{eqn:fco}
f & = & 1 \hspace{3.6cm}\mbox{if} \hspace{0.2cm}\mbox{HAS} \ge 1 \nonumber\\
     { } & = & 1 + m\cdot(\mbox{HAS}-1)  \hspace{1cm}\mbox{if} \hspace{0.2cm}\mbox{HAS} < 1
\end{eqnarray}
where HAS is the time after sunset measured in hours, $m$ is the same
for all bolometers and continuity at HAS$ = 1$ has been applied. We fit
the measured FCF$_i$/$<$FCF$_i$$>$ for all bolometers and all beam
maps with HAS $<$1 simultaneously to Equation \ref{eqn:fco} to
find $m=-0.115\pm0.002$~hr$^{-1}$.

\subsection{Calibration Error}
\label{cal:err}
For a given science observation with responsivity S$_i$ and extinction
e$^{-\tau_{\rm{eff}}}$ (measured from the bolometer dc-levels) the
calibrated timestream bolometer signals $\bar{I}_i$(t) are given by

\begin{equation}
\label{eqn:bsc}
\bar{I}_i(t) = {b_i(t) \cdot
<\mbox{FCF}_i> \cdot f \over S_i \cdot e^{-\tau_{\rm{eff}}}}
\end{equation}
where f is determined from the HAS that the observation took place and
the empirical formula derived above (Equation \ref{eqn:fco}). The
error on the calibrated bolometer signals is therefore equal to the
quadrature sum of the errors on all four factors in Equation
\ref{eqn:bsc} and is typically 6-13\% for the JCMT05B data.

\section{PERFORMANCE}
\label{sec:per} 
\subsection{Array Layout and System Efficiency}
\label{sec:performance}
\begin{figure}
\includegraphics[width=\hsize]{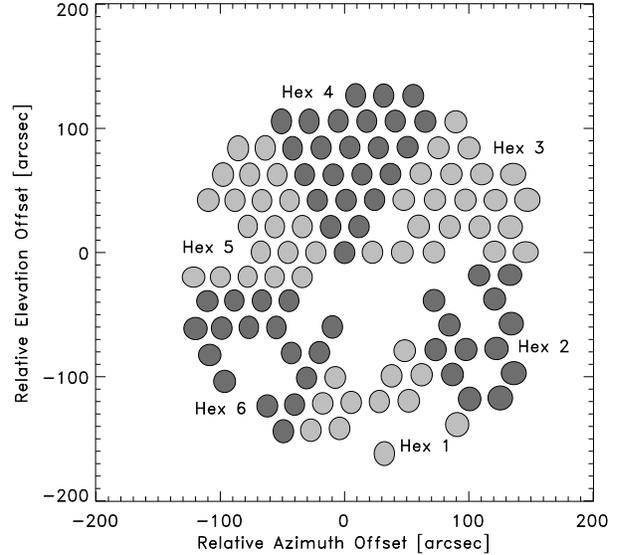}
\caption{\label{fig:arr} AzTEC's ``footprint'' on the sky at the
  JCMT05B run. The six hextants are labeled ``Hex'' 1-6. The alternate
  shading indicates which bolometers are located in each hextant. The
  size of the ellipse corresponds to the bolometer's FWHM, measured in the
  azimuth and elevation directions.}
\end{figure}

The array ``footprint'' on the sky for the JCMT05B run is shown is Figure
\ref{fig:arr}, centered on the reference bolometer and with the six
hextants labeled. We exclude all detectors that were not fully
operational during this run, most of which are located in hextants 1
and 2.  This left us with 107 operational pixels for the JCMT05B
season.  The majority of failures in the signal chains have recently
been traced to broken JFETs and these are expected to be repaired by
2009.

\begin{figure}
\includegraphics[width=\hsize]{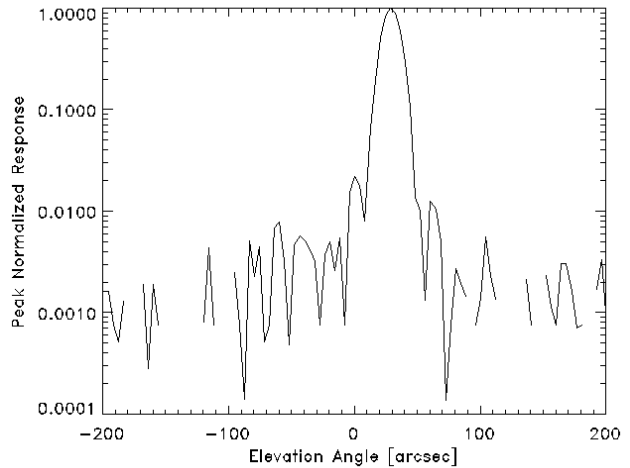}
\caption{\label{fig:beamcut} 
  A cut in the elevation direction through the beam response pattern of a 
  typical AzTEC pixel.  }
\end{figure}

Each bolometer's location on the array and its PSF are determined from
beam map observations as described in Section \ref{ovr:bmp}. The
positions and beam sizes displayed in Figure \ref{fig:arr} were
determined by averaging the measurements from all beam maps of Uranus
taken at the JCMT. The size of the ellipse is equal to the beam FWHM,
which is measured in the azimuth and elevation directions and is on
average $17^{\prime\prime} \pm 1^{\prime\prime}$ in azimuth and
$18^{\prime\prime} \pm 1^{\prime\prime}$ in elevation at the JCMT.  An
off-axis ellipsoidal mirror in the optics chain leads to the slight
elongation of the beam in the elevation direction. The array FOV is
roughly circular with a diameter of 5$^{\prime}$.  A constant azimuth cut
through a typical bolometer's beam map is shown in
Figure~\ref{fig:beamcut}.  The beams are nicely gaussian down to the
first sidelobe response at $\sim -20$dB.

Table~\ref{tab:optical_characteristics} lists the measured optical
characteristics of AzTEC in the 1.1~mm configuration for the JCMT05B
observations.  The effective throughput ($A\Omega\eta$) is
measured for each detector from each beammap of Uranus.  It is
calculated using Equation~\ref{eqn:fcf} along with AzTEC's measured
bandpass, $f(\nu)$ and the measured flux conversion factor of that
detector, FCF$_i$.  The value listed in
Table~\ref{tab:optical_characteristics} is an average over all working
detectors.

The optical efficiency $\eta$ of the system may be written as the
product of the telescope's efficiency, $\eta_\mathrm{tel}$, and the
coupling efficiency of the instrument $\eta_\mathrm{inst}$.  We
estimate the telescope emissivity, $\epsilon$, as the ratio between
the effective telescope temperature of Figure~\ref{fig:tau_v_atm}
(dashed line) and the true average temperature of the telescope
surface.  This emissivity estimate gives $\eta_\mathrm{tel}=
1-\epsilon=0.85$.  Removing this factor from the total throughput, we
find that the {\em instrument throughput},
$A\Omega\eta_\mathrm{inst}$, is 0.23~mm$^2$-sr.  Comparing this to the
idealized throughput of a single moded system at 1.1~mm wavelength
with the same telescope efficiency
($A\Omega\eta_\mathrm{tel}=\lambda^2\eta_\mathrm{tel}=0.99$) we find
the optical efficiency for the AzTEC radiometer to be 0.19.  Given
that the cold Lyot stop leads to a 38\% reduction in throughput, this
leaves an overall efficiency of 31\%.  This is very similar to the
optical efficiency measured for the Bolocam instrument
\citep{Glenn2003} which has a very similar optical design. 

\begin{table}
\begin{center}
\begin{tabular}{|l|c|}
\hline
Band center frequency   & 270.5~GHz \\
Effective bandwidth     & 49.0~GHz  \\
Effective throughput ($A\Omega\eta)$           & $0.2\pm 0.014~$mm$^2$ sr. \\
Beam FWHM (azimuth)     & 17\arcsec $\pm$ 1\arcsec \\
Beam FWHM (elevation)    & 18\arcsec $\pm$ 1\arcsec \\
\hline
\end{tabular}
\vspace{-.25cm}
\end{center}
\caption{AzTEC 1.1~mm optical parameters for the JCMT05B observations.  The effective bandwidth is calculated 
         assuming a flat-spectrum source.}
\label{tab:optical_characteristics}
\end{table}

\subsection{Noise and Sensitivity}

\begin{figure}
\includegraphics[width=0.9\hsize]{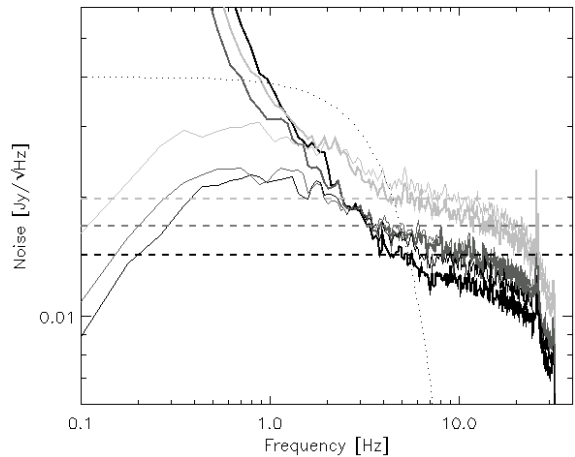}
\caption{The noise equivalent flux density (NEFD) for a typical
  bolometer.  In order of decreasing darkness, the three shades
  correspond to effective $\tau_{225}$ values of 0.11, 0.16 and 0.21.
  The thicker curves represent raw data while the thinner curves show
  data that have been cleaned using a principal component analysis
  (PCA).  The cleaned noise spectra are higher at some frequencies
  because they have been scaled to account for point-source
  attenuation due to cleaning.  The lower-opacity data benefits more
  from the common-mode removal effected by cleaning as well as by the
  reduced optical loading.  The dashed lines represent the
  corresponding NEFDs for a detector with targeted detector parameter
  values at our JCMT bias level along with an estimate of optical
  loading (details in the text).  The dotted curve indicates the
  approximate optical bandwidth of a point source (in arbitrary units)
  for a scan velocity of 180$^{\prime\prime}$/s.}
\label{fig:psds}
\end{figure}

Blank field observations done in raster-scan mode are used to estimate
detector noise and sensitivity.  Figure~\ref{fig:psds} shows the noise
equivalent flux density per beam (NEFD) for a typical detector in
three weather conditions at the JCMT.  The thicker curves show the raw
NEFDs while the thinner curves of the same shade show the improvement
due to a time stream cleaning method based on a principal component
analysis (PCA).  The low-frequency features, dominated by atmospheric
fluctuations, are not completely projected out by the PCA cleaning.
The flatter NEFD at higher frequencies that does not benefit from
cleaning can be attributed to the irreducible noise floor due to the
photon background limit (BLIP) and detector noise.

The three dashed lines indicate the thermodynamic noise limits for an
ideal AzTEC detector given the bias voltage, expected atmospheric
optical loading for three opacities, and optical loading from the
telescope (assuming a 15\% effective emissivity).  In each case the
actual high-frequency noise level is consistent with the detector and
loading model to within 10-20\%, well within our uncertainty of the
total optical loading on the detectors.  The nearly constant ratio
between the achieved and expected noise levels over varying weather
conditions indicates that the operationally convenient choice of using
a constant bias voltage for the entire observing run (see
section~\ref{sec:det}) had little if any negative impact on
sensitivity.  Table~\ref{tab:detector_noise} lists the expected detector operating parameters
and the breakdown of thermodynamic noise contributions for conditions
at the JCMT assuming $\tau_{225}=0.11$.  The actual post-cleaning
noise equivalent power measured for a typical detector is dominated by
residual low frequency noise due to the atmosphere and, based on our
point-source sensitivity, is equivalent to having a white noise level
of 1846 aW/$\sqrt \mathrm{Hz}$.
\begin{table}
\begin{center}
\begin{tabular}{c|c|c|c}
\hline
$T$ (mK )& $R$ (M$\Omega$) & $g$ (pW/K) & $S$ (V/W) \\
438 & 1.79 & 295 & $1.22\times10^8$ \\
\hline
\hline
\end{tabular}
\begin{tabular}{|l|c|}
Johnson noise   & 34 aW/$\sqrt{\rm{Hz}}$ \\
Phonon noise    & 44 aW/$\sqrt{\rm{Hz}}$ \\
Amplifier       & 58 aW/$\sqrt{\rm{Hz}}$ \\
Load resistor   & 4  aW/$\sqrt{\rm{Hz}}$ \\
Photon noise    & 102  aW/$\sqrt{\rm{Hz}}$ \\
{\bf Total:}    & {\bf  130 aW/$\sqrt{\rm{Hz}}$} \\
\hline
\end{tabular}
\vspace{-.25cm}
\end{center}
\caption{The nominal operational values for detector temperature
($T$), thermistor resistance ($R$), thermal-link conductance ($g$),
and bolometer responsivity ($S$) are given at the top.  They are
calculated according to the detector parameters and operating bias
given in section~\ref{sec:det} and the atmospheric loading expected
from a $\tau_{225}$ (opacity) of 0.11 given the optical
characteristics of Table~\ref{tab:optical_characteristics}.  Also
tabulated is the expected breakdown of thermodynamic noise
contributions.  They are calculated by using the above operating
parameters according to the equations of \citet{mather84a}.  Because
noise is quoted in terms of noise equivalent power (NEP) {\em absorbed
at the detector}, the optical parameters of
Table~\ref{tab:optical_characteristics} must be used for converting
these to noise equivalent surface brightness or flux density (as in
Table~\ref{tab:sensitivity}).}
\label{tab:detector_noise}
\end{table}

The increase in noise at low frequency highlights the importance of
scanning at high speed.  The dotted curve of Figure~\ref{fig:psds}
indicates the approximate optical bandpass of a detector at a scan
velocity of 180$^{\prime\prime}$/s, which is also the detector
response to a point source at that scan speed.  Throughout the course
of the observing run, we used scan speeds from 30$^{\prime\prime}$/s
to 270$^{\prime\prime}$/s based on balancing the opposing effects of
higher sensitivity and larger turn-around time fraction at higher scan
speeds given a map size, as explained in section~\ref{mod:ras}.  With
scans along the elevation direction, we do not see vibration-induced
noise increases within this range of velocities during the JCMT05B
observations.  We did measure excess noise near $\sim$2~Hz on
bad-weather nights at the JCMT, possibly due to wind-induced small
motions of optical loads and/or optics, like the JCMT's Gore-tex
cover.

The abrupt noise cut-off near 32~Hz is due to a digital filter that
conditions the demodulated bolometer signals for 64-Hz decimation.
The attenuation with frequency between 20 and 30~Hz is due to the
bolometer time constant.  The line at $\sim$25~Hz is likely a side
band caused by the third harmonic of AC power (or ``60~Hz'') mixed with
the 200~Hz demodulation waveform.  When analyzing raster-scan
observations we implement a second digital low-pass filter in software to avoid
aliasing this power back into the signal bandwidth.

\begin{figure}
\includegraphics[width=\hsize]{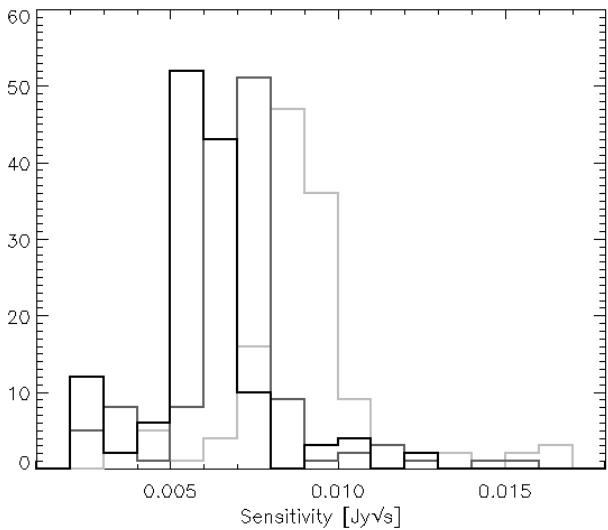}
\caption{Histogram of the ultimate sensitivity of working detectors for three 
different effective opacities using the higher frequency ``flat''
  noise level of raw data as seen in Figure~\ref{fig:psds}.}
\label{fig:sens}
\end{figure}

The ``flat'' NEFDs near 10~Hz represent a limiting white noise
level regardless of how effectively the low-frequency atmospheric
features can be removed using a particular cleaning method.
Therefore, in Figure~\ref{fig:sens} we have histogrammed these
ultimate sensitivities for the working detectors.  The particular
detector whose NEFD is shown in Figure~\ref{fig:psds} falls within one
of the better populated (tallest) bins of Figure~\ref{fig:sens} in all
three weather conditions.

\subsection{Mapping Speed and Instrument Sensitivity}
The best indicator of future performance and capability for an array
receiver is the instrument mapping speed.  Mapping speed is a metric
that can be summed linearly and simultaneously accounts for the
variations in detector sensitivities, the effectiveness of the
atmospheric cleaning algorithm, the individual optical efficiencies
achieved by each detector, and most importantly, the residual
correlations between detectors in the array.  We calculate the
effective mapping speed for our JCMT raster-scanned maps as
\begin{equation}
\label{eqn:mse}
M_\mathrm{em} = {N_\mathrm{det}\Omega_\mathrm{pix} \over
t_\mathrm{int}}
\sum\limits_{i=1}^{N_\mathrm{pix}} {1 \over
\sigma_i^2} ,
\end{equation}
where $\sigma_i$ represents the noise level of the $i$th pixel in a
map with $N_\mathrm{pix}$ pixels of solid angle $\Omega_\mathrm{pix}$.
$N_\mathrm{det}$ represents the number of functional detectors and $t_\mathrm{int}$
is the total integration time spent on a field.

\begin{figure}
\vspace{0.1in}
\includegraphics[width=\hsize]{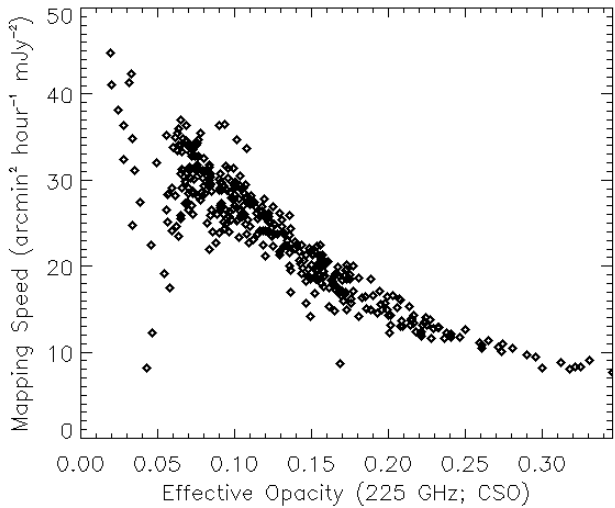}
\caption{
Empirical point-source mapping speeds for the AzTEC/JCMT system for a
variety of atmospheric opacities and using the current
cleaning/mapping software.  These mapping speeds were calculated according 
to Equation~\ref{eqn:mse} and do not include
overheads and mapping efficiencies that are specific to the observing
strategy employed.  Here $\sigma_i$ is taken as the raw (unfiltered) 
pixel noise scaled by the expected gain due to an optimal 
point-source filter after coaddition.  See text for details.}

\label{fig:msp} 
\end{figure}

Point-source mapping speeds achieved through raster scanning of the
JCMT05B observations of large blank fields are plotted in
Figure~\ref{fig:msp}.  Most overheads and mapping efficiencies are
ignored, making these idealized speeds generally applicable to
raster-scanned AzTEC maps at any observatory by scaling the mapping
speed values by the ratio of the telescope areas (assuming similar
telescope efficiencies).

It is generally favorable to apply a point-source filter \emph{after}
the coaddition of multiple raster-scanned maps due to the benefits of
cross-linking.  To estimate effective mapping speeds of the individual
observations presented in Figure~\ref{fig:msp}, $\sigma_i$ values of
the resulting map are estimated as the unfiltered pixel noise scaled
by the average reduction factor due to our optimum filter when applied
to a coadded map.

Mapping speed is correlated to the atmospheric conditions in two ways:
atmospheric loading (shot noise) and atmospheric stability.  The
latter gives rise to residual fluctuations that are inseparable from
astronomical signals which leads to much of the scatter in the mapping
speeds.  The high scatter in mapping speeds at the lowest opacities
are in-part due to errors in the estimation of the opacity.  There is
no significant evidence that the trend of mapping speed with opacity
breaks down for the lowest opacities.

Table~\ref{tab:sensitivity} gives the expected and achieved noise performance of the
instrument in a manner that allows one to compare ideal and achieved
detector sensitivities and mapping speeds in terms of flux-density.
In the table, the column of ``Projected'' sensitivities indicates the
sensitivity predicted from the bolometer model and the measured
optical loading at the JCMT.  The three columns of measured
sensitivities show the achieved sensitivities in the presence of
atmospheric noise in three cases: A) if ``perfect'' atmospheric noise
subtraction were possible as measured by the 10-Hz value of the time
stream detector noise of Figure~\ref{fig:psds}, B) with the achieved
atmospheric noise subtraction indicated by the thinner power
spectral density (PSD) curves of Figure~\ref{fig:psds} and the the
point-source response function (also shown as a dotted curve in that
figure), and C) as inferred from empirical mapping speed estimates
obtained with Equation~\ref{eqn:mse}.  We use the following
relationship between mapping speed and detector sensitivity ${\hat s}$
(in mJy$\sqrt{s}$) to calculate one where the other is known:
\be
\mathrm{MS} = \frac{3600 N_{\mathrm{det}}\Omega_b}{2{\hat s}^2}
\label{eqn:mst}
\ee
where $N_{\mathrm{det}}=107$ and
$\Omega_b=0.096$~arcminutes$^2$, the area under a
18\arcsec-FWHM 2-d gaussian.  Equation~\ref{eqn:mst} assumes the use
of a simple beam smoothing filter on maps.  The factor of $2 {\hat
s}^2$ in the denominator represents the square of ${\hat
s}_\mathrm{sm}$ ($= \sqrt 2 {\hat s}$), the appropriate sensitivity
post smoothing.

The degradation of sensitivity and mapping speed between columns 3 and
4 is believed to be due to non-idealities such as residual
bolometer-bolometer correlations which are not apparent from the
timestream PSDs.  We emphasize that column 4 of Table~\ref{tab:sensitivity}, or more
generally, Figure~\ref{fig:msp}, properly scaled by telescope area, is
the most accurate reference for planning observations with AzTEC.  The
values given in Table~\ref{tab:sensitivity} are only meant to be illustrative of where the
losses in sensitivity occur when beginning projections with raw
detector sensitivities.

\vspace{0.35cm}
\begin{table*}
\begin{center}
\begin{tabular}{rccccl}
\hline
      & Projected        & Measured          & Measured     & Measured    &  \\
      & (bolometer model) & (white PSD level) & (actual PSD) & (map space) &  \\    
      & JCMT (LMT)       & JCMT              & JCMT         & JCMT        &  \\ 
Detector Sensitivity& 10.02 (0.9) & 8.55  & {\bf 14.5}     & 26-28       & mJy$\sqrt{s}$ \\ 
Mapping Speed       & 184 (2013)  & 253  & 88             & {\bf 23-28} & arcmin$^2$/mJy$^2$/hr \\
\hline
\end{tabular}
\vspace{-.25cm}
\end{center}
\caption{Expected and achieved noise performance.  ``Projected'' is the
(white) timestream PSD prediction based on the bolometer model and
optical loading.  Values in parentheses are calculated for the LMT by
scaling according to telescope area assuming an effective LMT mirror
diameter of 43~m as (very conservatively) truncated by AzTEC's
Lyot stop.  ``Measured (white PSD level)'' is the noise level measured
from calibrated timestream PSDs at 10~Hz from the $\tau_{225}=0.11$
{\em thicker} curve of Figure~\ref{fig:psds}.  ``Measured (actual
PSD)'' is the inferred sensitivity based on the full cleaned time
stream PSD of the $\tau_{225}=0.11$ {\em thinner} curve of
Figure~\ref{fig:psds}.  The quoted sensitivity is calculated by
forming a weighted average of PSD/2, weighted by the square of the
point-source response function of figure~\ref{fig:psds}, and then
taking the square root.  ``Measured (map space)'' is the sensitivity
inferred from mapping speeds estimated with
Equation~\ref{eqn:mse}. Values given in bold are directly computed
from AzTEC data.  All other values are estimated as described in the
text.}
\label{tab:sensitivity}
\end{table*}

\section{FUTURE}
\label{sec:fut} 
AzTEC completed a successful 3-month scientific run on James
Clerk Maxwell Telescope from November 2005 to February 2006.
Reduction and analysis of the various data sets acquired during the
run are being completed.  AzTEC is now a facility instrument at the
10~m Atacama Submillimeter Telescope Experiment (ASTE; a Japanese ALMA
prototype telescope) for observations during the South American
winters of 2007 and 2008.  AzTEC will be delivered to the 50~m Large
Millimeter Telescope (LMT) in early 2009, where it will serve as a
first-light instrument at 2 and 1~mm wavelengths, and then become a
facility instrument for general use in all three of its wavebands.

\section*{acknowledgments}
We acknowledge the following people for their various and valuable
contributions to the success of the AzTEC instrument and/or AzTEC
observations: Itziar Aretxaga, Cara Battersby, Yuxi Chen, Iain
Coulson, Gary Davis, Simon Doyle, Daniel Ferrusca, Douglas Haig,
Salman Hameed, David Hughes, John Karakla, James Lowenthal, Gary
Wallace, and Miguel Valazquez. SK and YK were supported in part by
Korea Science \& Engineering Foundation (KOSEF) under a cooperative
agreement with the Astrophysical Research Center of the Structure and
Evolution of the Cosmos (ARCSEC).  This work was supported in part by
NSF Grant \#0540852.

\bibliography{references}

\end{document}